\documentclass{article}

\usepackage{arxiv}

\usepackage[utf8]{inputenc} 
\usepackage[T1]{fontenc}    
\usepackage{hyperref}       
\usepackage{url}            
\usepackage{booktabs}       
\usepackage{amsfonts}       
\usepackage{nicefrac}       
\usepackage{microtype}      
\usepackage{lipsum}
\usepackage{graphicx}
\usepackage[square,numbers]{natbib}
\usepackage{doi}

\usepackage{amsmath}
\usepackage{bm}
\DeclareMathOperator{\atantwo}{atan2}
\DeclareMathOperator{\atan}{atan}

\DeclareMathOperator{\asin}{asin}
\DeclareMathOperator{\vex}{vex}
\DeclareMathOperator{\tr}{tr}

\newtheorem{prop}{Proposition}
\usepackage{amssymb}
\usepackage{mathtools}

\graphicspath{ {./images/} }

\title{Control of convertible UAV with vectorized thrust}

\author{
 Tomas Lopes de Oliveira \\
 I3S (University Cote d’Azur, CNRS)\\ 
  Nice, France 06160 \\
  \texttt{tlopes@i3s.unice.fr} \\
   \And
 Tarek Hamel \\
I3S (University Cote d’Azur, CNRS)\\ 
 and Insitut Universitaire de France\\
 Nice, France 06160 \\
 \texttt{thamel@i3s.unice.fr} \\
  \And
 Claude Samson \\
 I3S (University Cote d’Azur, CNRS)\\ 
 Nice, France 06160 \\
 \texttt{csamson@i3s.unice.fr} \\
 \And
 André Anglade \\
 I3S (University Cote d’Azur, CNRS)\\ 
 Nice, France 06160 \\
 \texttt{aanglade@i3s.unice.fr} \\
}

\renewcommand{\rhead}{Author accepted version}
\renewcommand{\undertitle}{Author accepted version \thanks{\hspace{.2em}To appear in IFAC-PapersOnLine 2023 \\
\textcopyright 2023 the authors. This work has been accepted to IFAC for publication under a Creative Commons Licence CC-BY-NC-ND.}}
\renewcommand{\shorttitle}{Control of convertible UAV with vectorized thrust}

\begin{document}
\maketitle

\makeatletter

\begin{abstract}                
This paper is an addition to an article previously published by three of the authors that addresses the control of 
convertible fixed-wing aircraft with vectorized thrust.
The control solution here developed extends the one presented in the former paper by complementing it with a strategy for the hovering flight and a control policy to handle the transition phase between low and high airspeed. An estimator of the air velocity required in all flight phases is also proposed.
Realistic simulation results on a tri-tilt rotor Unmanned Aerial Vehicle (UAV) illustrate and assess the methodology.

\end{abstract}


\section{Introduction}
\begin{figure}
    \begin{center}
    \includegraphics[width=10cm]{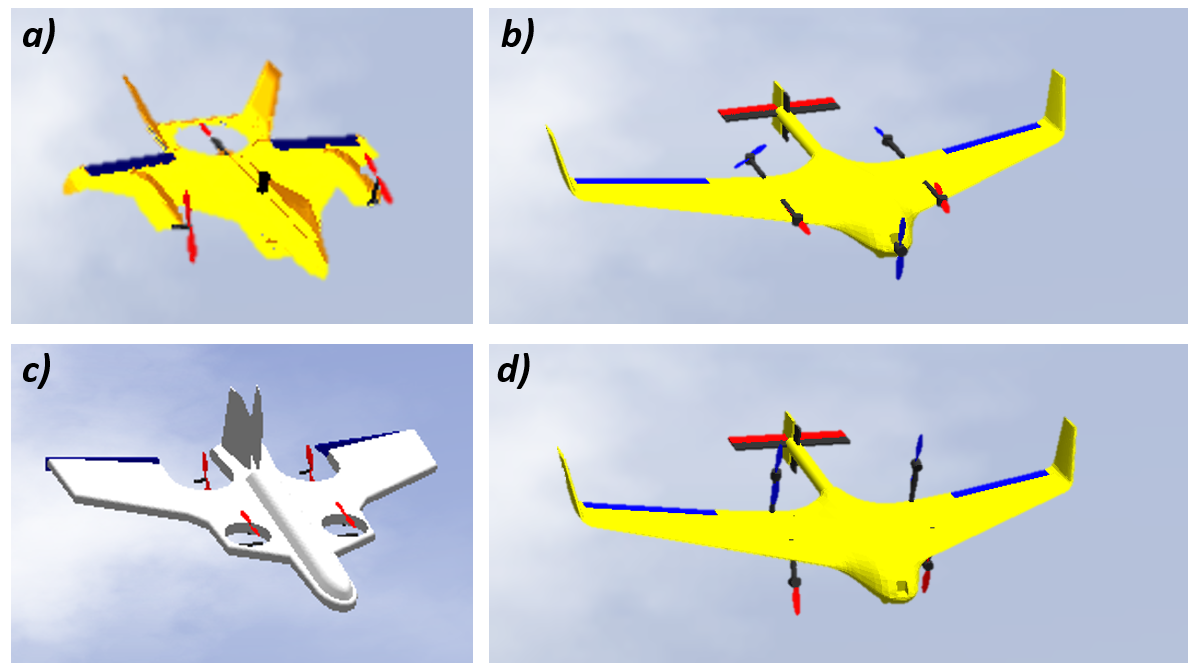}    
    \caption{Examples of convertible fixed-wing aircraft with vectorized thrust:
    in (a), the Eflite Convergence VTOL with one rear fixed motor and two front ones tilting independently;
    in (b), a quad-airplane with four fixed lifting motors plus one fixed front motor; in
    (c), the FV-31 Cypher VTOL with two front fixed motors and two rear ones tilting together;
    in (d), a quadtilt-airplane with four motors tilting together.}
    \label{fig: drones}
    \end{center}
\end{figure}

Convertible fixed-wing aircraft with thrust vectoring capabilities are aerial vehicles for which transition from hover to cruising flight is achieved by tilting the thrust vector with respect to (w.r.t.) the main body. The versatility of these vehicles has historically motivated the interest of major aircraft manufacturers, even though only a few prototypes have reached an operational phase (e.g., Bell V-22 Osprey and Lockheed Martin F-35 Lightning II). The expanding market of small Unmanned Aerial Vehicles (UAVs) has renewed this interest, as exemplified by many representatives of this class of vehicles now rendered available to the general public. Popular configurations are inspired by the combination of fixed-wing aircraft and quadrotors, like the examples represented in Fig \ref{fig: drones}.

Controlling convertible aerial vehicles necessarily involves a transition between low-speed stages (hover) and high speeds ones (cruising flight of airplanes). However, most of the solutions proposed in the literature fail to address within a unified control framework the monitoring of the delicate transition phase itself, and of the accompanying tilting of the vectorized thrust.
Indeed, the proposed transition strategy commonly relies on a scheduling policy in the horizontal plane when the aircraft accelerates (resp. decelerates) from hovering to cruising mode (resp. cruising to hovering mode). The tilt angle is typically chosen, independently of the aerodynamic forces acting on the aircraft, as a function of the aircraft's airspeed \cite{liu2017control,ducard2021review,flores2014nonlinear}. Experimental validations supporting this strategy, although not numerous, can be found in \cite{flores2022longitudinal}. In \cite{ciopcia2017quad}, the authors advocate for the minimization of the thrust intensity via the choice of the tilt angle, which they consider a perspective worth to be investigated. This possibility has subsequently been addressed in \cite{anglade2019automatic} with the development of a general control framework applicable to a large class of aerial vehicles, including fixed-wing convertible aircraft.
The proposed control solution applies to most airspeed and flight conditions, but there remain singular conditions for which it is ill-defined. Hovering at zero, or at reduced, airspeed is one of them due to the ill-definition of the angle of attack in this mode. This issue is pointed out in \cite{willis2021nonlinear}, where a controller that stabilizes the desired pitch independently of the desired yaw is proposed to handle this case. The present paper proposes another solution that extends the applicability of the generic controller in \cite{anglade2019automatic} via complements that encompass both hovering and the transition phase.Another addition concerns the (approximate) estimation of the three-dimensional air velocity vector, whose knowledge is important to reduce the control sensitivity to (unknown) aerial perturbations, in the case where only the airspeed in the longitudinal body direction is measured (with a pitot tube, for instance).

The article is organized as follows: Section \ref{sec: recalls on control} recalls the control methodology reported in \cite{anglade2019automatic}. Section \ref{complement} proposes complements to encompass the hovering flight and the transition between low and high speed. Section \ref{sec: application} applies the control framework to a light tri-tilt rotor UAV, with simulation results presented in Section \ref{sec: experiment}.
\section{Control model}
\subsection{Notation}
\begin{figure}
    \begin{center}
    \includegraphics[width=10cm]{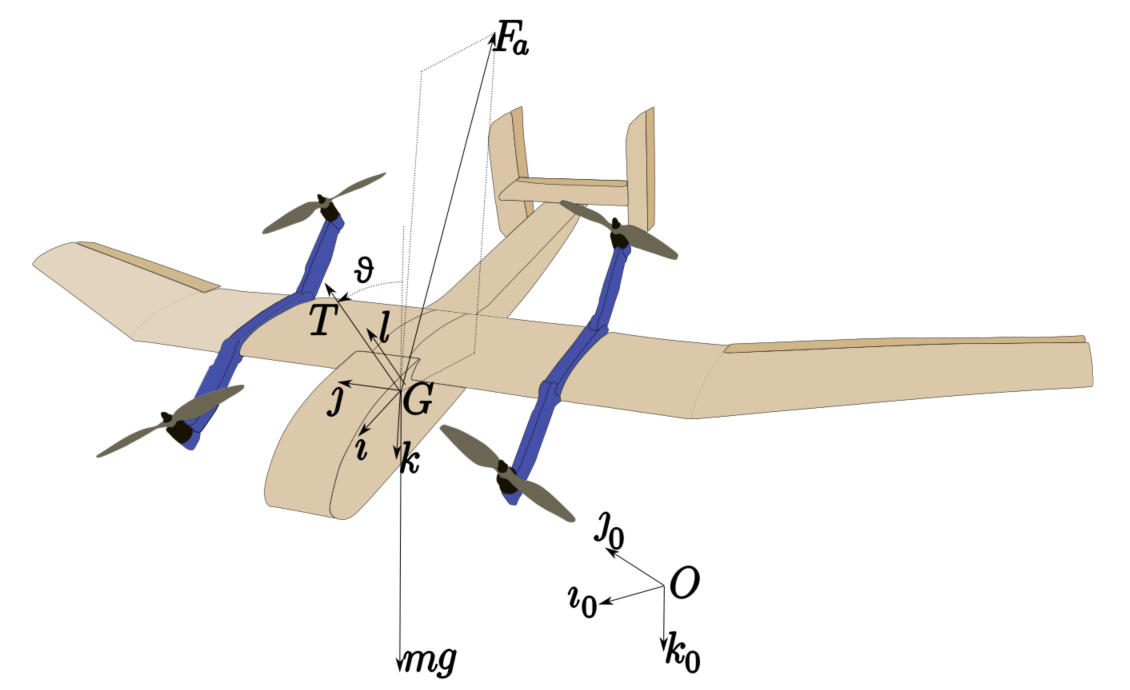}    
    \caption{Frames and forces}
    \label{fig:forces and frames}
    \end{center}
\end{figure}

\begin{itemize}
    \item $\bm{E}^3$ denotes the 3D Euclidean space, and $\mathcal{I} = \{ O, \bm{\imath}_0, \bm{\jmath}_0, \bm{k}_0 \}$ is a right-handed inertial frame with fixed origin $O$ and the vector $\bm{k}_0$ pointing vertically downward.
    \item Vectors in $\bm{E}^3$  are written with bold letters. Inner and cross products in $\bm{E}^3$ and $\mathbb{R}^3$ are denoted by the symbols $\cdot$ and $\times$ respectively. Ordinary letters are used for real vectors of coordinates, and the $\mathit{ith}$. component of a vector $x$ is denoted as $x_i$.
    \item  When $x \in \mathbb{R}^n$ (resp. $\bm{x} \in \bm{E}^3$), $|x|$ (resp. $|\bm{x}|$) denotes the Euclidean norm of $x$ (resp. $\bm{x}$). If $x$ is a vector of coordinates of $\bm{x}$ in some frame, then $|x| = |\bm{x}|$.
    \item $\Pi_{\bm{u}}$ is the projection operator on the plane orthogonal to the vector $\bm{u}$.
    \item $\mathbb{P}_{a}$ is the anti-symmetric operator in square matrix space given by $\mathbb{P}_{a}(H) = \frac{1}{2}(H-H^T)$ and $(\cdot)_{\times}$ denotes the skew-symmetric matrix associated with the cross product, i.e., $\forall a,b \in \mathbb{R}^3$  $a_{\times}b = a \times b$.
    \item $m$ is the aircraft mass. $J$ is the inertia matrix calculated w.r.t. the aircraft center of mass (CoM), denoted by $G$.
    \item  $\mathcal{B} = \{ G,\bm{\imath}, \bm{\jmath}, \bm{k}\}$ is the chosen aircraft-fixed frame, with $\bm{\imath}$ and $\bm{\jmath}$ parallel to the so-called zero-lift plane of the aircraft. In this paper, it is assumed that this latter plane is not affected by thrust direction changes. This implies, in particular, that the aircraft is designed so as to minimize aerodynamic interference between the propulsion system and the vehicle body. The vector $\bm{\imath}$ (resp  $\bm{\jmath}$) is aligned with the longitudinal (resp. lateral) axis of the aircraft (see Fig. \ref{fig:forces and frames}).
    \item $\bm{l} = \sin(\vartheta)\bm{\imath} - \cos(\vartheta)\bm{k}$ is the thrust unit direction, with $\vartheta$ the tilting thrust angle.
    \item The matrix $R \in SO(3)$ that encodes the rotation of the frame $\mathcal{B}$ w.r.t. $\mathcal{I}$ represents the vehicle's attitude. The column vectors of $R$ correspond to the vectors of coordinates of $\bm{\imath}$, $\bm{\jmath}$, $\bm{k}$ expressed in the basis of $\mathcal{I}$.
    \item $\bm{\omega}$ is the angular velocity of $\mathcal{B}$ w.r.t. $\mathcal{I}$, i.e.
    \begin{equation}
    \frac{d}{dt}(\bm{\imath}, \bm{\jmath}, \bm{k}) = \bm{\omega} \times (\bm{\imath}, \bm{\jmath}, \bm{k}).
    \end{equation}
    The vector of coordinates of $\bm{\omega}$ in the body-fixed frame $\mathcal{B}$ is denoted $\omega$.
    \item $\bm{p}$, $\bm{v}$ and $\dot{\bm{v}}$ are, respectively, the position, the velocity and the acceleration of the G w.r.t to $\mathcal{I}$. Position and velocity are related by the kinematic equation
    \begin{equation}\label{eq: kinematic model}
    \dot{\bm{p}} = \bm{v}.
    \end{equation}
    \item $\bm{g} = g_0 \bm{k_0}$ is the gravity vector.
    \item $\bm{v_w}$ is the ambient wind velocity w.r.t. $\mathcal{I}$, which is assumed bounded, with first and second-time derivatives also bounded.
    \item $\bm{v}_a = \bm{v} - \bm{v}_w$ is the aircraft air-velocity. The vector of coordinates of $\bm{v}_a$ in $\mathcal{B}$ is denoted $v_a$.
    \item The direction of $\bm{v}_a$ in $\mathcal{B}$ is expressed as a function of two angles $\alpha$ and $\beta$, such that
    \begin{equation}\label{eq: va in body frame}
        \bm{v}_a = |v_a|(\cos\alpha(\cos\beta\bm{\imath} + \sin\beta\bm{\jmath}) + \sin{\alpha}\bm{k}).
    \end{equation}
     where $\alpha = \arcsin(v_{a,3}/|v_a|)$ denotes the angle of attack, and $\beta = \arctan(v_{a,2}/|v_{a,1}|)$ denotes the side-slip angle.
\end{itemize}
In the remainder of the paper, measurements or estimates of the aircraft attitude, angular velocity, inertial position and velocity are assumed to be available. Controlling motor tilt angles at desired values does not pose a particular problem for small aerial vehicles, and is thus omitted.

\subsection{Generic system equations of a convertible aircraft with vectorized thrust}
This section recalls the system equations of a generic convertible fixed-wing, as proposed in \cite{anglade2019automatic}, for which complementary thrust tilting capabilities achieve the transition from hover to cruising flight. The independent control inputs here considered are a control torque vector $\bm{\Gamma}$ capable of modifying the vehicle's orientation at will and a thrust vector $\bm{T}:=T\bm{l}$, with $T$ denoting the thrust magnitude. The aircraft is subjected to the aerodynamic force $\bm{F}_a$ given by the following expression:
\begin{equation}\label{eq: aerodynamic force}
    \bm{F}_a = -(c_0(\bm{v}_a \cdot \bm{\imath})\bm{\imath} + \bar{c}_0(\bm{v}_a \cdot \bm{k})\bm{k})|v_a| + (\bm{v}_a \cdot \bm{\jmath})\bm{O}(\bm{v}_a)
\end{equation}
where $c_0$ and $\bar{c}_0$ denote positive numbers, and $\bm{O}(\bm{v}_a)$ denotes an Euclidean vector-valued function such that the ratio $\frac{|\bm{O}(\bm{v}_a)|}{|\bm{v}_a|}$ is bounded. More details about this model can be found in \cite{anglade2019automatic} and \cite{kai2019unified}.

The system of equations describing the aircraft dynamics is given by the combination of \eqref{eq: kinematic model} and the classical Newton-Euler equations:
\begin{equation}\label{eq: system equation}
 m\dot{\bm{v}} = m\bm{g} + \bm{F}_{a} + T(\sin (\vartheta) \bm{\imath}  - \cos(\vartheta) \bm{k}  ),
\end{equation}
\begin{equation}\label{eq: system angular velocity equation}
    J\dot{\omega} = -\omega \times J\omega + \Gamma \ (+ \text{parasitic terms}),
\end{equation}
 with $\Gamma$ the vector coordinates of $\bm{\Gamma}$ in the body-fixed frame.

\section{Control design}\label{sec: recalls on control}
The previous section shows that the system involves only five control inputs for six degrees of freedom which forbids complete and decoupled control of the vehicle's position and attitude. The control problem is addressed via the formalism of primary and secondary objectives by extending the solution previously derived in the fixed thrust-direction case \cite{kai2017nonlinear}. The primary control objective is related to the vehicle's longitudinal motion. Let  $\bm{\xi}$ denote a 3-dimensional real vector specified later on.
Define
\begin{equation}\label{eq: def of a,d and e}
    \begin{array}{lll}
         & \bm{a}: = m(\bm{\xi} - \bm{g}), \\
         & \bm{d}: = \bm{a} + c_0|v_a|\bm{v}_a,\\
         & \bm{e}: = \bm{a} + \bar{c}_0|v_a|\bm{v}_a.
    \end{array}
    \end{equation}
Substituting \eqref{eq: aerodynamic force} along with \eqref{eq: def of a,d and e} in \eqref{eq: system equation}, it yields:
\begin{equation}\label{eq: dynamics with pos objective}
    \begin{array}{cc}
        &  m\dot{\bm{v}} = m \bm{\xi} + (T\sin(\vartheta)- \bm{d} \cdot \bm{\imath}) \bm{\imath}\\
        &- (T\cos(\vartheta) + \bm{e} \cdot \bm{k}) \bm{k} - (\bm{a} \cdot \bm{\jmath} )\bm{\jmath} - (\bm{v_a} \cdot \bm{\jmath})\bm{O}(\bm{v_a}).
    \end{array}
\end{equation}
Denote $\bar{\mathcal{B}} = (G,\bar{\bm{\imath}},\bar{\bm{\jmath}},\bar{\bm{k}})$ as a desired body-fixed frame so that $\bar{\bm{\jmath}}$ is orthogonal to $\bm{a}$ and $\bm{v}_a$:
\begin{equation}\label{eq: j constraint}
    \bar{\bm{\jmath}} = \frac{\bm{v}_a \times \bm{a}}{|\bm{v}_a \times \bm{a}|},
\end{equation}
and set $ T\sin(\vartheta) = \bm{d} \cdot \bar{\bm{\imath}}$ and $T\cos(\vartheta) = -\bm{e} \cdot \bar{\bm{k}}$ so that
\begin{equation}\label{eq: constraints 1}
    \begin{array}{cc}
        & \tan(\vartheta) = -\frac{\bm{d} \cdot \bar{\bm{\imath}}}{\bm{e} \cdot \bar{\bm{k}}},\\
        & T = \sin(\vartheta)(\bm{d} \cdot \bar{\bm{\imath}}) -\cos(\vartheta)(\bm{e} \cdot \bar{\bm{k}}).
    \end{array}
\end{equation}
One can rewrite \eqref{eq: dynamics with pos objective}, as follows:
\begin{equation}\label{eq: frame convergence to pos obj}
    \begin{array}{cc}
         & m\dot{\bm{v}} = m \bm{\xi} - [\bm{d} \cdot (\bm{\imath} - \bar{\bm{\imath}}) ] \bm{\imath} - [\bm{e} \cdot (\bm{k} - \bar{\bm{k}})] \bm{k} + \\
         & \quad \quad \quad + [\bm{a}\cdot (\bm{\jmath} - \bar{\bm{\jmath}})]\bm{\jmath} - [\bm{v_a}\cdot (\bm{\jmath} - \bar{\bm{\jmath}})]\bm{O}(\bm{v_a}).
    \end{array}
\end{equation}
Hence, if $T$ and $|v_a|$ are bounded, and if $(\bm{\imath},\bm{\jmath},\bm{k})$ asymptotically converges to $(\bar{\bm{\imath}},\bar{\bm{\jmath}},\bar{\bm{k}})$, then:
\begin{equation*}
    \dot{\bm{v}} = \bm{\xi} + o(t), \ \  \lim_{t \rightarrow \infty} o(t) = 0.
\end{equation*}
The position control problem is brought back to a simple linear control one with $\bm{\xi}$ the new control input. Examples for the design of $\bm{\xi}$ in relation to commonly considered control objectives (stabilization, tracking, and path-following) can be found in \cite{anglade2019automatic}.

Note that the above control framework does not impose any desired rotation around the axis $\bar{\bm{\jmath}}$. A secondary control objective thus concerns the choice of this rotation.
\ In \cite{anglade2019automatic}, two possibilities are proposed:
\begin{itemize}
    \item[1)] determine the angle of attack that minimizes the thrust intensity $T$:
    \begin{equation}\label{alphah1}
        \alpha = 0.5 \atantwo(|\bm{v}_a \times \bm{a}|,\bm{a} \cdot \bm{v}_a+0.5(c_0+\bar{c}_0)|v_a|^3),
    \end{equation}
    \item[2)] choose the tilt angle $\vartheta$ as a function of the air velocity and then determine the corresponding angle of attack that satisfy constraints \eqref{eq: j constraint} and \eqref{eq: constraints 1}:
    \begin{equation}\label{alphah2}
        \alpha = \atantwo\Biggl(\frac{\sin(\vartheta)|\bm{v}_a \times \bm{a}| -\cos(\vartheta)(\bm{a} \cdot \bm{v}_a + c_0|v_a|^3 ) }{\cos(\vartheta)|\bm{v}_a \times \bm{a}| +\sin(\vartheta)(\bm{a} \cdot \bm{v}_a + \bar{c}_0|v_a|^3 ) }\Biggr).
    \end{equation}
\end{itemize}
Once the desired angle of attack $\alpha$ is defined, the frame vectors $\bar{\bm{\imath}}$ and $\bar{\bm{k}}$ are given by
\begin{subequations}\label{eq: previous frame i bar and k bar}
    \begin{align}
         &    \bar{\bm{\imath}} = \cos (\alpha)\frac{\bm{v}_a }{|v_a|} +\sin (\alpha) \Big( \bar{\bm{\jmath}} \times \frac{\bm{v}_a}{|v_a|} \Big), \\
         &  \bar{\bm{k}} = \bar{\bm{\imath}} \times \bar{\bm{\jmath}} = \sin (\alpha)\frac{\bm{v}_a}{|v_a|} -\cos (\alpha) \Big( \bar{\bm{\jmath}} \times \frac{\bm{v}_a }{|v_a|} \Big).\label{eq: previous k bar}
    \end{align}
\end{subequations}
 Let $\bar{\bm{\omega}}$ denote the angular velocity of the desired frame $\bar{\mathcal{B}}$, i.e. $\bar{\bm{\omega}} = \bar{\bm{\omega}}_{\bar{\bm{k}}} + (\bar{\bm{k}} \cdot \bar{\bm{\omega}}_{\bar{\bm{\imath}}})\bar{\bm{k}}$, with $\bar{\bm{\omega}}_{\bar{\bm{\imath}}}: = \bar{\bm{\imath}} \times \dot{\bar{\bm{\imath}}}$ and $\bar{\bm{\omega}}_{\bar{\bm{k}}}: = \bar{\bm{k}} \times \dot{\bar{\bm{k}}}$.
One shows that the following angular velocity vector
\begin{equation}\label{eq: desired angular velocity}
    \bm{\omega}^* = \bar{\bm{\omega}} + k_{\bm{\imath}}(t)(\bm{\imath} \times \bar{\bm{\imath}}) +  k_{\bm{\jmath}}(t)(\bm{\jmath} \times \bar{\bm{\jmath}}) +  k_{\bm{k}}(t)(\bm{k} \times \bar{\bm{k}}),
\end{equation}
with  $k_{\bm{\imath}}(t), k_{\bm{\jmath}}(t), k_{\bm{k}}(t)>\epsilon>0$,
exponentially stabilizes $\mathcal{B}=\bar{\mathcal{B}}$. See \cite{kai2017nonlinear} for the proof.

From there, one is left with the determination of a control torque $\bm{\Gamma}$ yielding the closed-loop exponential stabilization at zero of the angular velocity error $\omega - \omega^*$. From  \eqref{eq: system angular velocity equation}, with torque parasite terms neglected, a possible choice is (see \cite{anglade2019automatic}):
\begin{equation*}
    \Gamma = -k_{\gamma}J(\omega - \omega^*) + \omega \times J\omega^* + J\dot{\omega}^*,
\end{equation*}
where $k_{\gamma}>0$ denotes a control gain. In practice, one may consider a simple high gain control $\Gamma = -k_{\gamma}J(\omega - \omega^*)$, with $k_{\gamma}$ a diagonal matrix with positive elements.

This generic control framework can be applied to a large class of thrust-tilting convertible aircraft for which the airspeed is larger than some threshold. However, it is not directly applicable to hovering flight because the desired orientation involves the expression of $\bar{\bm{\jmath}}$, and of the angle of attack $\alpha$, that are not defined at zero air velocity. 
A way of circumventing this difficulty in practice is presented next.

\section{Extension of the control design to hovering flight}\label{complement}
\subsection{Low-speed attitude control}\label{subsec: low-speed}
The expression \eqref{eq: j constraint} reveals that $\bar{\mathcal{B}}$ is not well-defined when
either $|v_a|$ or $|\bm{a}|$ are equal to zero, or when $\bm{v}_a$ is aligned with $\bm{a}$. In practice, the problem extends to the case when the vehicle hovers or moves slowly in the absence of wind.
However, it remains possible to impose a desired pitch angle for the vehicle thanks to the thrust tilting ability. Let $\theta_d$ denote this angle, chosen in the interval of $[0,\frac{\pi}{4})$ for instance. Consider the following set of vectors $ \bm{\imath}_d$, $\bar{\bm{k}}$ and $\bm{\jmath}_d$:
\begin{subequations}\label{eq: low speed frame}
\begin{align}
     &   \bm{\imath}_d = \cos (\theta_d) \frac{\Pi_{\bm{k}_0}(\bm{\imath})}{|\Pi_{\bm{k}_0}(\bm{\imath})|} - \sin(\theta_d) \bm{\bm{k}_0} \label{subeq: i bar low speed frame definition},\\
     & \bar{\bm{k}} = -\frac{\Pi_{\bm{\imath}_d}(\bm{a})}{|\Pi_{\bm{\imath}_d}(\bm{a})|} \label{eq: k bar in pseudo frame},\\
     & \bm{\jmath}_d = \frac{\bar{\bm{k}} \times \bm{v}_a}{\epsilon + |\bar{\bm{k}} \times \bm{v}_a|},
     \end{align}
\end{subequations}
where $\epsilon$ is a small positive number that prevents a division by zero. In practice, one may replace $\bm{\imath}_d$ by a filtered version of this vector that ensures the boundedness of its first derivative. The unit vector $\bar{\bm{k}}$ is, by definition, orthogonal to $\bm{\imath}_d$ and $\bm{\jmath}_d$.
Moreover, for $|v_a|$ larger than a threshold $\sigma>0$, setting 
$\bar{\bm{\jmath}}:=\frac{\bm{\jmath}_d}{|\bm{\jmath}_d|}$ and $\bar{\bm{\imath}}:=\frac{\bar{\bm{k}}\times \bar{\bm{\jmath}}}{|\bar{\bm{k}}\times \bar{\bm{\jmath}|}}$ yields a frame $(G, \bar{\bm{\imath}},\bar{\bm{\jmath}},\bar{\bm{k}})$
that can be used as a body frame. In the absence of sideslip, the corresponding desired angle of attack $\alpha$ is equal to
\begin{equation}\label{alpha_l}
\alpha = \asin\Big(\frac{\bm{v}_a}{|v_a|} \cdot \bar{\bm{k}}\Big).
\end{equation}
Consider now the following desired angular velocity:
\begin{align}\label{eq: desired angular velocity2}
    \bm{\omega}^* &= \bar{\bm{\omega}} + k_{\bm{\jmath}}(t)(\bm{\jmath} \times \bm{\jmath}_d) +  k_{\bm{k}}(t)(\bm{k} \times \bar{\bm{k}}),
\end{align}
where  $k_{\bm{\jmath}}(t), k_{\bm{k}}(t)>\epsilon>0$ and $\bar{\bm{\omega}} = \bar{\bm{\omega}}_{\bar{\bm{k}}} + (\bar{\bm{k}} \cdot \bar{\bm{\omega}}_{\bar{\bm{\jmath}}})\bar{\bm{k}}$, with $\bar{\bm{\omega}}_{\bar{\bm{\jmath}}}: = \bar{\bm{\jmath}}\times \dot{\bar{\bm{\jmath}}}$ and $\bar{\bm{\omega}}_{\bar{\bm{k}}}: = \bar{\bm{k}} \times \dot{\bar{\bm{k}}}$.

Using the fact that two unit vectors $\bm{\jmath}_d$ and $\bar{\bm{k}}$
are enough to characterize the orientation of the desired frame $\bar{\mathcal{B}}$, one shows the following result:
\begin{prop} \label{prop1}
If $|v_a|\geq \sigma>0, \; \forall t \geq 0$, then the virtual control $ \omega^*$ in \eqref{eq: desired angular velocity2} almost globally exponentially stabilizes $\mathcal{B}=\bar{\mathcal{B}}$.  If $|v_a|=0, \; \forall t$, then $\omega^*$  almost globally exponentially stabilizes $\bm{k}=\bar{\bm{k}}$.
\end{prop}
A proof is given in the appendix.
The low-speed control description is completed by specifying the thrust vector with tilt and intensity given respectively, by
\begin{equation}\label{eq: constraints 1 for low frame}
    \begin{array}{cc}
        & \tan(\vartheta) = -\frac{\bm{d} \cdot \bm{\imath}_d}{\bm{e} \cdot \bar{\bm{k}}},\\
        & T = \sin(\vartheta)(\bm{d} \cdot \bm{\imath}_d) -\cos(\vartheta)(\bm{e} \cdot \bar{\bm{k}}).
    \end{array}
\end{equation}

The transition from low hovering speed to high cruising speed involves the air velocity. Thus, before presenting the transition control strategy, the issue of measuring, or estimating, this velocity is addressed in the next subsection.
\subsection{Air velocity estimation}\label{subsec: airvelestim}
The computation of the desired body frame $(\bar{\bm{\imath}},\bar{\bm{\jmath}},\bar{\bm{k}})$  and of $\vartheta$ and $T$ depend on the measurement, or estimation, of $\bm{v}_a$. For small UAVs, direct measurements of the inertial velocity $\bm{v}$ are available. Direct measurements of $\bm{v}_a$ are not, in part because size and weight constraints prevent small aircraft from being equipped with AoA vanes, or with more than one pitot tube, thus restricting the measurement of $\bm{v}_a$ to a single component of this vector (e.g. $v_{a_1}$). Typically,  these measurements are unreliable when $v_{a_1}$ is under a minimum threshold.
In this case, and in the absence of a better alternative, we propose to apply the control solution reported in subsection \ref{subsec: low-speed} with the estimation of $\bm{v}_a$ taken equal to the null vector.
Larger measured values of $v_{a_1}$ can be used to estimate the other two components, $v_{a_2}$ and $v_{a_3}$.

One method is to assume that the wind blows essentially horizontally. Suppose also that the sideslip velocity is small. Then, one can approximate $v_{a_3}$ by

\begin{equation}
    \hat{v}_{a_3} = \frac{\bm{k}_0 \cdot (\bm{v}-v_{a_1}\bm{\imath})}{\bm{k}_0 \cdot \bm{k}}.
\end{equation}
By further assuming that the wind velocity is approximately constant, one has $\dot{\bm{v}}_a = \dot{\bm{v}}$ and the derivative of $(\bm{v}_a \cdot \bm{\jmath})$ is $\dot{v}_{a_2} = \dot{\bm{v}} \cdot \bm{\jmath} + \bm{v}_a \cdot (\bm{\omega} \times \bm{\jmath})$. Hence, by \eqref{eq: aerodynamic force} and \eqref{eq: system equation},
\begin{equation}\label{eq: lateral derivative}
    \dot{v}_{a_2} = (\bm{g} \cdot \bm{\jmath}) + \bm{O}(\bm{v}_a)v_{a_2}  + (v_{a_3} \omega_1 - v_{a_1}\omega_3).
\end{equation}
Notice that when the airspeed is large, the dissipative term $\bm{O}(\bm{v}_a)v_{a_2}$ may, thanks to the rear vertical
stabilizer, produce a dominant torque that maintains the sideslip velocity $v_{a_2}$ small. An estimator of $v_{a_2}$ that mimics \eqref{eq: lateral derivative} is given by
\begin{equation}
    \dot{\hat{v}}_{a_2} = (\bm{g} \cdot \bm{\jmath}) + (\hat{v}_{a_3} \omega_1 - v_{a_1}\omega_3) - k_{v_{a_2}}\hat{v}_{a_2},
\end{equation}
where $k_{v_{a_2}}$ is a positive stabilizing gain to be adjusted in relation to the specific aircraft properties.
\subsection{Transition control strategy}
The proposed control strategy for the transition between low and high airspeed involves two domains of the flight envelope. The first one is the low airspeed domain for which $|v_a|<\sigma_m$, with $\sigma_m>0$ denoting a threshold.  Below this threshold, one applies the control strategy described in section \ref{subsec: low-speed} for hovering and low velocities.

The second domain is characterized by $|v_a| \geq \sigma_m$ and
encompasses the transition between low and high airspeed. Denote $\alpha_l$ the angle of attack calculated according to \eqref{alpha_l}, and $\alpha_h$ the optimal angle of attack given by \eqref{alphah1} or \eqref{alphah2}. Define an interval $(\sigma_m,\sigma_M)$ delimiting the air velocities for which the transition occurs, and define the desired angle of attack $\alpha$ as
\begin{equation}\label{eq: general frame}
    \alpha = \lambda(|v_a|)\alpha_l + (1-\lambda(|v_a|))\alpha_h,
\end{equation}
where the function $\lambda(|v_a|)$ is a non-increasing smooth function on $\mathbb{R}^+$ such that $\lambda(x)=1$ when $x \leq \sigma_m$, and $\lambda(x) = 0$ when $ x \geq \sigma_M$. This new angle of attack is
introduced to allow for a smooth transition and used to determine the unit vectors of the desired body frame according to \eqref{eq: j constraint} and \eqref{eq: previous frame i bar and k bar}. These vectors are in turn used to calculate the desired body angular velocity
$\bm{\omega}^*$ according to \eqref{eq: desired angular velocity}.

\section{Adaptation to the Eflite UAV}\label{sec: application}

\begin{figure}
    \begin{center}
    \includegraphics[width=10cm]{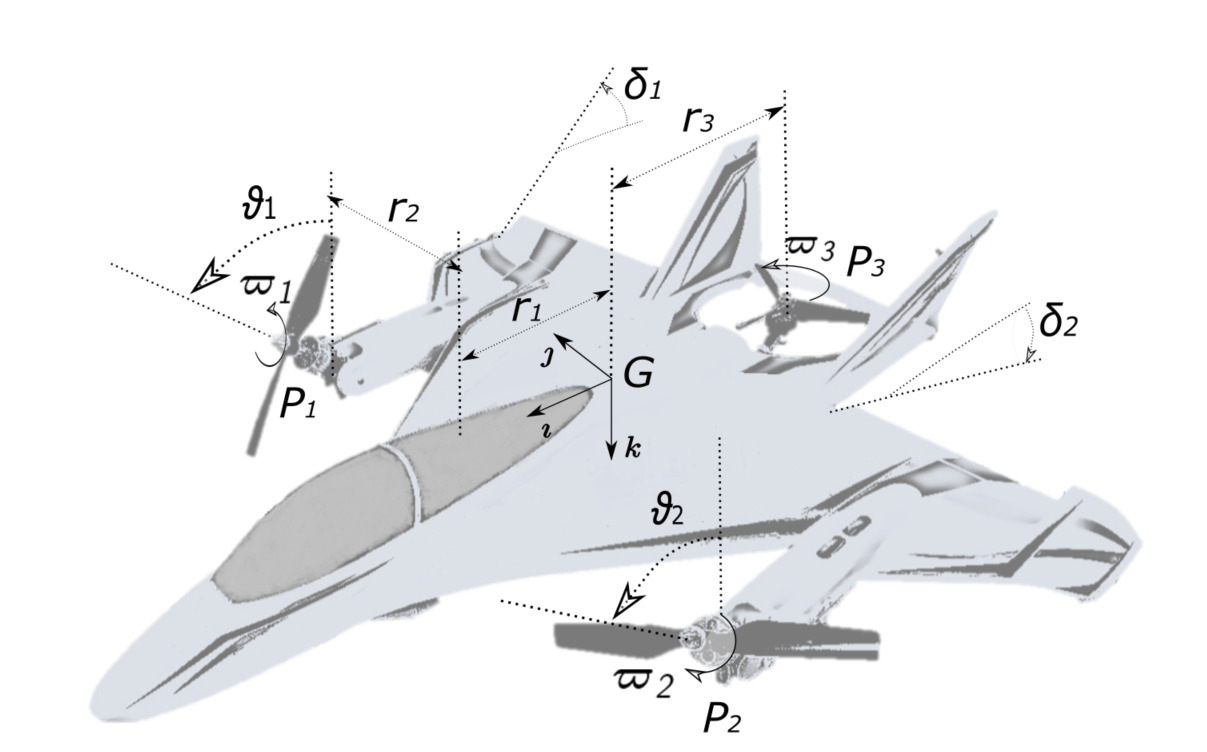}    
    \caption{Eflite Convergence configuration}
    \label{fig: EfliteConvergence}
    \end{center}
\end{figure}

The Eflite Convergence UAV (see Fig. \ref{fig: EfliteConvergence}) is a convertible fixed-wing UAV with two control surfaces and three rotors (two of which are equipped with an independent tilting mechanism).
Therefore, the control inputs are the deflection angles $\delta_j$ ($j = 1,2$), rotor-tilt angles $\vartheta_m$ ($m = 1,2$) and rotation speeds of the rotors $\varpi_n$ ($n = 1,2,3$) as depicted in Fig. \ref{fig: EfliteConvergence}.
The rotors and the control surfaces are in charge of producing the control torque $\Gamma$ specified in section \ref{sec: recalls on control}. This torque can in turn be decomposed into:
\begin{equation*}
    \Gamma = \Gamma_a + \Gamma_m,
\end{equation*}
where $\Gamma_a$ is the aerodynamic torque input produced by the control surfaces, and  $\Gamma_m$ is the torque input produced by the rotors. Control surfaces are efficient only at high airspeed, and rotors are more efficient at low airspeed. This suggests in \cite{anglade2019automatic} a torque transition of the form:
\begin{equation*}
    \Gamma_a =(1-\bar{\lambda}(|v_a|))\Gamma, \ \ \ \Gamma_m=\bar{\lambda}(|v_a|)\Gamma,
\end{equation*}
with $\bar{\lambda}$ a non-increasing smooth function on $\mathbb{R}^+$ such that $\bar{\lambda}(x) = 1$ for $0 \leq x \leq \delta^*$, and $\bar{\lambda}(x) = 0$ for $x > \delta^* + r$, with $\delta^*>0$ and $r>0$ characterizing the chosen velocity interval for the torque transition.

The next step is to compute the actuator inputs that produce ($\bm{T}$,$\Gamma_m$,$\Gamma_a$). The determination of the angles $\delta = [\delta_1,\delta_2]^T$ is directly given by (see \cite{anglade2019automatic}):
\begin{equation*}
    \delta = \frac{1}{|v_a|^2}A\Gamma_a.
\end{equation*}
The matrix $A$ depends on the size of the control surfaces and their location w.r.t. to $G$. As for the thrust vector, it is computed from the desired inputs $(T,\vartheta)$, and the desired control torque $\Gamma_m$. For the Eflite Convergence UAV, one determines the following system of equation, also reported in \cite{willis2021nonlinear}:
\begin{equation}\label{eq: explicit motor actuation system}
    \begin{bmatrix}
        T\sin(\vartheta)\\
        T\cos(\vartheta)\\
        \Gamma_m
    \end{bmatrix} =
        D
    \begin{bmatrix}
        \mu_{12} \varpi_1^2\cos(\vartheta_1)\\
        \mu_{12} \varpi_2^2\cos(\vartheta_2)\\
        \mu_{12} \varpi_1^2\sin(\vartheta_1)\\
        \mu_{12} \varpi_2^2\sin(\vartheta_2)\\
        \mu_{3}\varpi_3^2\\
    \end{bmatrix},
\end{equation}
where $\mu_{12}$ and $\mu_{3}$ and are positive constants and D is the following invertible matrix:
\begin{equation*}
    D =
    \begin{bmatrix}
        0 & 0 & 1 & 1 & 0\\
        1 & 1 & 0 & 0 & 1\\
        -r_2 & r_2 & \nu_{12} & -\nu_{12} & 0\\
        r_1 &  r_1 & 0 & 0 & -r_3\\
        -\nu_{12} & \nu_{12} & -r_2 & r_2 & -\nu_{3}\\
    \end{bmatrix}
\end{equation*}
with $r_i$ ($i=1,2,3$), $\nu_{12}$ and $\nu_{3}$ positive constants. In view of \eqref{eq: explicit motor actuation system}, by defining
\begin{equation*}
    X := D^{-1}
    \begin{bmatrix}
       T\sin(\vartheta) & T\cos(\vartheta) & \Gamma_m^T
    \end{bmatrix}^T,
\end{equation*}
one computes $\varpi_m$ and $\vartheta_n$ via:
\begin{equation*}
   \begin{array}{llll}
    \mu_{12} \varpi_1^2 = \sqrt{X_1^2 + X_3^2}, \ 
    \mu_{12} \varpi_2^2 = \sqrt{X_2^2 + X_4^2},\\
    \mu_{3} \varpi_3^2 = X_5, \
    \vartheta_1 = \atan \Big( \frac{X_3}{X_1} \Big), \
    \vartheta_2 = \atan  \Big( \frac{X_4}{X_2} \Big),
   \end{array}
\end{equation*}
with $X_1,X_2,X_3,X_4$ and $X_5$ the components of $X$.

In practice, a certain number of limitations have to be taken into account, like saturations for $\varpi_m$ ($\omega_{\min} \leq \varpi_m \leq \omega_{\max}$, $m=1,2,3$) and for $\vartheta_n$ ($-\frac{\pi}{15} \leq \vartheta_n \leq \frac{\pi}{2}$, $n=1,2$). These issues are also of practical importance, but they are not addressed in this paper for lack of space.

\section{Software in the loop simulation}\label{sec: experiment}
This section reports software-in-the-loop (SITL) simulation results illustrating the performance of the control approach in a realistic scenario. The control law was implemented as an additional module of the open-source PX4 autopilot flight controller (\cite{meier2015px4}). The virtual environment for the simulation is the Gazebo robot simulator (https://github.com/PX4/sitl\_gazebo), along with open-source plugins based on the RotorS Project that enable the interfacing of the simulator with the controller, and with the ground station software. A gazebo customized model of the aerodynamic forces acting on the aircraft is used, based on the superposition of elementary disks covering the aircraft body and wings as proposed in \cite{hua2014novel}.

SITL simulations have been performed on the four fixed-wing convertible UAV models shown in Fig. \ref{fig: drones}. The models are equipped with the common sensor suite comprising a GNSS, an IMU, and a pitot tube. An Extended Kalman Filter (EKF), already embedded in PX4 firmware, provides the inertial velocity, position, and attitude estimations. In addition, the estimator described in section \ref{subsec: airvelestim} was implemented for the air velocity estimation.

A model of the Eflite Convergence aircraft is here used for the reported simulation results. The chosen flight mission consists in following an inclined circular path with a desired inertial speed incrementally increasing from $\sigma_m = 3m/s$ to $\sigma_M = 9m/s$, thus forcing the vehicle to cross the transition phase while turning. The control surfaces are used to produce a torque action for velocities larger that $\delta^*=7m/s$. As for the secondary control objective, the desired pitch angle $\theta_d = 0$ is chosen for low speed. Thrust minimization is used to determine the angle of attack at high speed.

\begin{figure}
    \begin{center}
    \includegraphics[width=8cm]{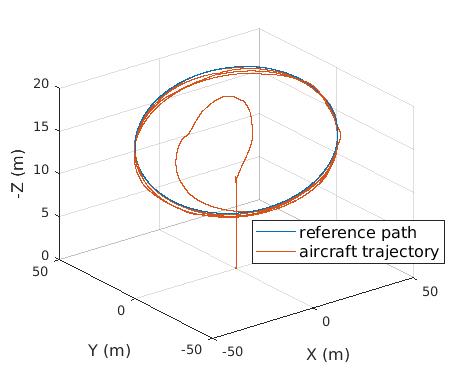}    
    \caption{Comparison between the desired inclined circular path and aircraft's trajectory.}
    \label{fig: traj}
    \end{center}
\end{figure}

After performing a short takeoff maneuver, with the auxiliary control input $\bm{\xi}$ (see section \ref{sec: recalls on control}) determined in the trajectory tracking mode, $\bm{\xi}$ is switched to a path following flight mode to execute an inclined circular path (see Fig. \ref{fig: traj}). The expression of $\bm{\xi}$ is then given by (see \cite{anglade2019automatic}):
\begin{equation*}
    \bm{\xi} = \xi_v \bm{h} + |\bm v|(\bar{\bm{\omega}}_h \times \bm{h}),
\end{equation*}
where $\bm{h} := \bm{v}/|\bm{v}|$, is the so-called heading direction (\cite{kai2019unified}). The term $\xi_v$ is the part of the feedback control in charge of making $|\bm{v}|$ converge to the desired speed $v^*$. The term $\bar{\bm{\omega}}_h$ is the part in charge of aligning $\bm{h}$ with the desired heading $\bm{h}^*$ and of making the distance to the chosen path converge to zero. Example functions for $\bar{\bm{\omega}}_h$ and $\xi_v$ are given in \cite{kai2019unified}.

\begin{figure}
    \begin{center}
    \includegraphics[width=11cm]{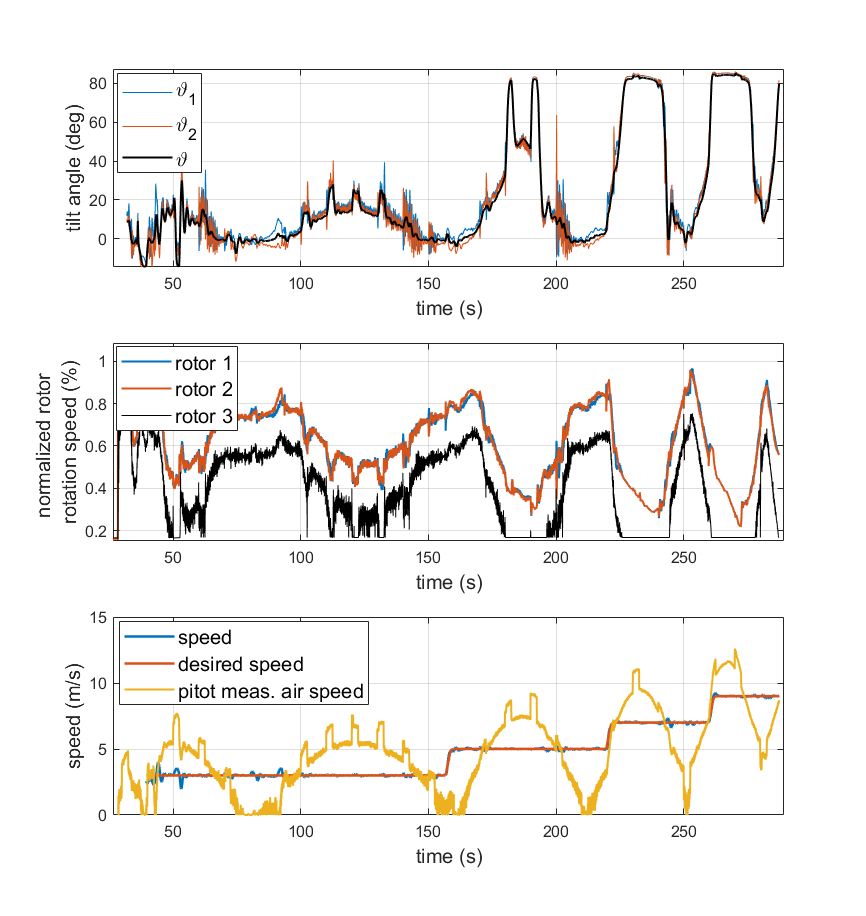}    
    \caption{On the top, tilt angle commands for the two motor tilts ($\vartheta_1,\vartheta_2$) and the thrust vector tilt ($\vartheta$). In the middle, normalized rotor rotation speed for the three motors. On the bottom, comparison between aircraft inertial and air speeds and the desired inertial speed.}
    \label{fig: motor_tilt_speed}
    \end{center}
\end{figure}

Simulations are performed in a challenging windy environment for small UAVs. The wind is chosen in the $\bm{\imath}_0$ direction with its magnitude given by the superposition of two components: a constant $3m/s$ wind speed, modified by $2m/s$ wind gusts that are applied during $2s$ every $10s$ during the flight. Figures \ref{fig: traj} and \ref{fig: motor_tilt_speed} (on the bottom) show the practical convergence of the vehicle position to the reference path at the desired inertial speed. Figure \ref{fig: motor_tilt_speed} shows the variation of the tilt angles and the rotor commands. As expected, the thrust tilting angle varies from zero degree (for low airspeed) to 86 degrees (for high airspeed), while the thrust magnitude decreases in relation to the creation of lift when the airspeed increases.

\section{Concluding Remarks}
This paper presents practical complements to a control framework for convertible vehicles with vectorized thrust, as proposed by the authors in \cite{anglade2019automatic}.
The control methodology has been tested extensively in software-in-the-loop simulation on several convertible aircraft models. Results on a realistic model of the Eflite Convergence UAV, and in a challenging windy environment, are reported to demonstrate the performance of the approach. Successful initial experiments reported in: 
\url{https://youtu.be/egjDarEUQJw}
 constitute an enticement to
now perform intensive experimentation on various convertible vehicles with vectorized thrust.


\bibliography{ifacconf}

\begin{thebibliography}{11}
\providecommand{\natexlab}[1]{#1}
\providecommand{\url}[1]{\texttt{#1}}
\expandafter\ifx\csname urlstyle\endcsname\relax
  \providecommand{\doi}[1]{doi: #1}\else
  \providecommand{\doi}{doi: \begingroup \urlstyle{rm}\Url}\fi

\bibitem[Anglade et~al.(2019)Anglade, Kai, Hamel, and
  Samson]{anglade2019automatic}
A.~Anglade, J.-M. Kai, T.~Hamel, and C.~Samson.
\newblock Automatic control of convertible fixed-wing drones with vectorized
  thrust.
\newblock In \emph{2019 IEEE 58th Conference on Decision and Control (CDC)},
  pages 5880--5887. IEEE, 2019.

\bibitem[Ciopcia and Szczepa{\'n}ski(2017)]{ciopcia2017quad}
M.~Ciopcia and C.~Szczepa{\'n}ski.
\newblock Quad-tiltrotor--modelling and control.
\newblock \emph{Journal of Marine Engineering \& Technology}, 16\penalty0
  (4):\penalty0 331--336, 2017.

\bibitem[Ducard and Allenspach(2021)]{ducard2021review}
G.~J. Ducard and M.~Allenspach.
\newblock Review of designs and flight control techniques of hybrid and
  convertible vtol uavs.
\newblock \emph{Aerospace Science and Technology}, 118:\penalty0 107035, 2021.

\bibitem[Flores(2022)]{flores2022longitudinal}
G.~Flores.
\newblock Longitudinal modeling and control for the convertible unmanned aerial
  vehicle: Theory and experiments.
\newblock \emph{ISA transactions}, 122:\penalty0 312--335, 2022.

\bibitem[Flores-Colunga and Lozano-Leal(2014)]{flores2014nonlinear}
G.~R. Flores-Colunga and R.~Lozano-Leal.
\newblock A nonlinear control law for hover to level flight for the quad
  tilt-rotor uav.
\newblock \emph{IFAC Proceedings Volumes}, 47\penalty0 (3):\penalty0
  11055--11059, 2014.

\bibitem[Hua et~al.(2014)Hua, Pucci, Hamel, Morin, and Samson]{hua2014novel}
M.-D. Hua, D.~Pucci, T.~Hamel, P.~Morin, and C.~Samson.
\newblock A novel approach to the automatic control of scale model airplanes.
\newblock In \emph{53rd IEEE Conference on Decision and Control}, pages
  805--812. IEEE, 2014.

\bibitem[Kai et~al.(2017)Kai, Hamel, and Samson]{kai2017nonlinear}
J.-M. Kai, T.~Hamel, and C.~Samson.
\newblock A nonlinear approach to the control of a disc-shaped aircraft.
\newblock In \emph{2017 IEEE 56th Annual Conference on Decision and Control
  (CDC)}, pages 2750--2755. IEEE, 2017.

\bibitem[Kai et~al.(2019)Kai, Hamel, and Samson]{kai2019unified}
J.-M. Kai, T.~Hamel, and C.~Samson.
\newblock A unified approach to fixed-wing aircraft path following guidance and
  control.
\newblock \emph{Automatica}, 108:\penalty0 108491, 2019.

\bibitem[Liu et~al.(2017)Liu, He, Yang, and Han]{liu2017control}
Z.~Liu, Y.~He, L.~Yang, and J.~Han.
\newblock Control techniques of tilt rotor unmanned aerial vehicle systems: A
  review.
\newblock \emph{Chinese Journal of Aeronautics}, 30\penalty0 (1):\penalty0
  135--148, 2017.

\bibitem[Meier et~al.(2015)Meier, Honegger, and Pollefeys]{meier2015px4}
L.~Meier, D.~Honegger, and M.~Pollefeys.
\newblock Px4: A node-based multithreaded open source robotics framework for
  deeply embedded platforms.
\newblock In \emph{2015 IEEE international conference on robotics and
  automation (ICRA)}, pages 6235--6240. IEEE, 2015.

\bibitem[Willis and Beard(2021)]{willis2021nonlinear}
J.~B. Willis and R.~W. Beard.
\newblock Nonlinear trajectory tracking control for winged evtol uavs.
\newblock In \emph{2021 American Control Conference (ACC)}, pages 1687--1692.
  IEEE, 2021.

\end{thebibliography}








\appendix    
\section{Proof of Proposition \ref{prop1}}

In what follows, $R$ ( resp. $\bar{R}$) denotes the rotation matrix associated with the frame $\mathcal{B}$ (resp. $\mathcal{\bar{B}}$).
Consider the Lyapunov function candidate defined by
\begin{equation*}
E= (1 - \imath^T\bar{\imath}) + (1 - \jmath^T\bar{\jmath}) + (1 - k^T \bar{k})
\end{equation*}

    Recall that for two matrices $A$ and $B \in \mathbb{R}^{n\times n}$ such that $A$ is a skew-symmetric matrix, one has $tr(AB) = tr(BA)=tr(\mathbb{P}_{a}(B)A)$, and that for any $a$ and $b \in \mathbb{R}^n$, $a^T b=tr(ab^T)$.  By defining $\tilde{R} := R\bar{R}^T$ such that $\imath=\tilde{R}\bar{\imath}$ (resp. $\jmath=\tilde{R}\bar{\jmath}$ and $k=\tilde{R}\bar{k}$), one verifies that $E = tr(I - \tilde{R})$, with $I$ denoting the identity matrix. Then, the first time derivative of $E$ is given by
    \begin{equation*}
        \dot{E}  = -\tr[(\omega - \bar{\omega})_{\times} \tilde{R} + \bar{\omega}_{\times}\tilde{R} - \tilde{R}\bar{\omega}_{\times}].
    \end{equation*}
    Setting $\omega = \omega^*$, and noticing that $tr(\bar{\omega}_{\times}\tilde{R}) = tr(\tilde{R}\bar{\omega}_{\times}) $,
    \begin{equation}
        \begin{array}{ll}
            &         \dot{E}  = -\tr\{[k_{\bm{k}}(k \times \bar{k})  +k_{\bm{\jmath}}(|v_a|)(\jmath \times \bar{\jmath})]_{\times }\tilde{R}\} \\ \nonumber
            & =  -\tr\{[k_{\bm{{k}}}(\tilde{R}\bar{k} \times \bar{k})  +k_{\bm{\jmath}}(|v_a|)(\tilde{R}\bar{\jmath} \times \bar{\jmath})]_{\times }\tilde{R}\}
        \end{array}
    \end{equation}
Using the fact,  that $(a \times b)_{\times} = b a^T - ab^T$, for $a$ and $b \in \mathbb{R}^3$, one verifies that:
    \begin{equation} \label{dE}
        \begin{array}{llc}
            & \dot{E}  = -\tr[k_{\bm{k}}(\bar{k}\bar{k}^T-\tilde{R}\bar{k}\bar{k}^T\tilde{R}) + \\
            & \ \ \ \ \ \ \ \ \ \ \ \ \ \ \ \ \ \ \ \ \ \ \ \ + k_{\bm{\jmath}}(|v_a|)(\bar{\jmath}\bar{\jmath}^T-\tilde{R}\bar{\jmath}\bar{\jmath}^T\tilde{R})]\\
            & = -k_{\bm{k}}\bar{k}^T(I-\tilde{R}\tilde{R})\bar{k}- k_{\bm{\jmath}}(|v_a|)\bar{\jmath}^T(I-\tilde{R}\tilde{R})\bar{\jmath}
        \end{array}
    \end{equation}
 By using in \eqref{dE} the following identity for any vector $a \in \mathbb{R}^3$ :
 \begin{equation}\label{identity}
      a^T\mathbb{P}_a(\tilde{R})\mathbb{P}_a(\tilde{R})a =  -\frac{1}{2}a^T(I-\tilde{R}\tilde{R})a,
 \end{equation}
one gets:
\begin{equation*}
    \dot{E} = -k_{\bm{k}}\bar{k}^T\mathbb{P}_a(\tilde{R})\mathbb{P}_a(\tilde{R})\bar{k}- k_{\bm{\jmath}}(|v_a|)\bar{\jmath}^T\mathbb{P}_a(\tilde{R})\mathbb{P}_a(\tilde{R})\bar{\jmath}.
\end{equation*}
Finally, recalling that for any $a$ and $b \in \mathbb{R}^3$, $a_{\times} b = -b_{\times} a$, one obtains
\begin{equation}\label{dotE}
\begin{array}{cc}
     &  \dot{E}= -\vex(\mathbb{P}_a(\tilde{R}))^T Q\vex(\mathbb{P}_a(\tilde{R}))\\
     & \leq - \lambda_{\min}(Q)||\mathbb{P}_a(\tilde{R})||^2,
\end{array}
\end{equation}
where $Q: = k_{\bm{k}} \Pi_{\bar{\bm{k}}} + k_{\bm{\jmath}}(|v_a|)\Pi_{\bar{\bm{\jmath}}} $ is a uniformly definite positive matrix as long as  $|v_a|>\sigma >0$. Almost global exponential stability of $\tilde{R}=I$, and thus of $\mathcal{B}=\bar{\mathcal{B}}$, follows directly. 

As for the case $|v_a|=0$, $\forall t$, \eqref{dotE} shows that $\dot{E}$ is decreasing as long as $\bm{k}$ is not colinear with $\bar{\bm k}$.
To further prove that the equilibrium $\bm{k}$ converges exponentially to $\bar{\bm k}$, it suffices to first verify that, by setting $\omega = \omega^*$: 
\begin{equation}\label{eq: d(1-kk)}
\frac{d}{d t}(1-k^T\bar{k})=-k_{\bm k}|k \times \bar{k}|^2
\end{equation}
By defining $\tilde{\theta}$ as the angle between  $\bm{k}$ and $\bar{\bm k}$, relation \eqref{eq: d(1-kk)} is equivalent to:
\begin{equation*}
\frac{d}{d t}\left( \sin^2\left(\frac{\tilde{\theta}}{2}\right) \right)=- 
2 k_{\bm k}\cos^2 \left(\frac{\tilde{\theta}}{2}\right)\sin^2\left(\frac{\tilde{\theta}}{2}\right)
\end{equation*}
Almost global exponential stability of $\tilde{\theta}=0$, and thus of $\bar{\bm k}=\bm k$, follows directly.

\end{document}